\documentstyle[12pt,cite,epsf]{article}
\begin{document}

\setlength{\parindent}{-0.5cm}
\vspace{-1.2cm}
\setlength{\parindent}{0.9cm}
\vspace{1.2cm}
\begin{center}

\noindent{\bf NEAR-FIELD INDUCED \\
FIR JOSEPHSON-DETECTION \\
BY c-AXIS-ORIENTED YBa$_2$Cu$_3$O$_{7-\delta}$-FILMS}

\vspace{7mm}

{\bf E.~Zepezauer$^1$, A.~Kalbeck$^1$, S.~D.~Ganichev$^{1,2}$,
W.~Korzenietz$^1$, and W.~Prettl$^1$}

{\small \it

\vspace{0.6cm}
$^1$Institut f\"ur Experimentelle und Angewandte Physik, \\
Universit\"at Regensburg, 93040 Regensburg, Germany\\
$^2$A. F. Ioffe Physicotechnical Institute of the RAS,\\
St. Petersburg 194021, Russia
}

\end{center}

\vspace{2.3cm}
\begin{center}
\section*{Abstract}
\end{center}

A novel approach to intrinsic Josephson-detection of far infrared
radiation is reported utilizing near-zone field effects at
electric contacts on c-axis oriented
YBa$_2$Cu$_3$O$_{7-\delta}$ films. While only a bolometric
signal was observed focusing the radiation far off the contacts
on c-axis normal
films, irradiating the edge of contacts
yielded an almost wavelength independent fast signal showing
the characteristic intensity dependence of
Josephson-detection. The signal is attributed to a c-axis
parallel component of the electric radiation field being
generated in the near-zone field of diffraction at the
metallic contact structures.

\vspace{1cm}
{\bf Key words:} near field, far infrared, Josephson effect,\\
YBa$_2$Cu$_3$O$_{7-\delta}$.

\clearpage
\section{Introduction}

Thin granular superconducting films of high-T$_c$
superconductors, biased close to their critical currents,
have been shown to be fast and sensitive detectors for the far
infrared (FIR) and millimeter wave
regime \cite{Leung87Optical,Huggard91Fast,
Schneider92Spectral,Huggard94Wavelength,Huber95Square}.
These films consist of stoichiometric
crystallites, each of which is coupled to its neighbors by weak
links.
The films are thus comprised of a random array of
Josephson junctions, the critical currents of which are depressed by
below gap radiation. Measured responsivities are
unsurprisingly found to vary considerably between samples, due to the
reliance of the detectivity on the properties of the random array.
Inherent Josephson junctions have also been found in
small single crystals of Bi- and Tl-based high-$T_c$ superconducting
materials \cite{Kleiner92Intrinsic}.
In this case, in contrast to granular films, a regular microscopic
array of junctions is formed along the c-axis by the quasi two-dimensionality of the material.
However, for efficient FIR radiation detection with
these small crystallites, a component of the electric field of the
radiation must be applied along the c-axis of the superconductor.
This is only realizable by using delicate antennas
to contact a small crystal with $\mu$m dimensions.
A considerable improvement in FIR Josephson detection
including difference frequency mixing has been achieved
by using thin film structures grown with a controllable
misalignment between the c-axis and the substrate surface
normal\cite{Huber96,Huber97}.
The coherent growth of such films on appropriately oriented substrates
has been established by the observation of a tilt angle dependent
lateral thermoelectric effect in the normal state \cite{Zeuner95Fast}.
In such films normal incidence irradiation  applies an electric field along
the c-axis.

Here we report on a new approach to FIR-Josephson detection
using c-axis oriented epitaxial grown YBa$_2$Cu$_3$O$_{7-\delta}$ films
which are of much better
quality than those prepared with a tilt angle to the c-axis.
The necessary component of the electric field of the FIR
radiation normal to the film surface is achieved by making
use  of the optical near-zone field effect
of diffraction on suitable electrode structures on the
sample surface.
Similar diffraction generated near-zone
field enhancement of electron tunneling has been previously
observed applying high-power FIR radiation on n-GaAs
tunnel Schottky diodes \cite{Gan96Schottky} and between metallic contacts and
$\delta$-doped 2DEG in n-GaAs 20~nm below the
surface\cite{Kotelnikov96}.

\section{Experimental}

Epitaxial YBa$_2$Cu$_3$O$_{7-\delta}$-layers have been prepared by
UV-laser evaporation on (100)LaAlO$_3$-substrates. The thickness
of the films was 300~nm. The samples were oxygen
depleted by thermally forced oxygen diffusion in Ar
atmosphere at 600~K for several hours.

\begin{figure}[ht!]

\centerline{\mbox{\epsfxsize=9cm \epsfbox{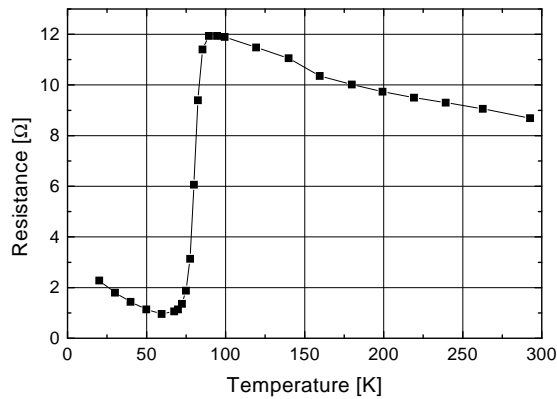}}}
\caption{\it{Sample resistance as function of temperature showing
semiconductor like behavior below T$_c$.}}

\label{RT}
\end{figure}

In Fig.~\ref{RT} a plot of resistance as a function of temperature is
shown for the samples used here with T$_c$ around T~=~80~K.
Due to oxygen deficiency the samples show semiconductor-like
behavior below and above T$_c$, the slope of the resistance
versus temperature is negative away from T$_c$. This allows
to distinguish easily between Josephson- and bolometric
signal as they differ in polarity.

On top of the films diffracting electrode structures were
deposited by sputtering large non-transparent parallel Au stripe-contacts
with various spacing widths around 100~$\mu$m in the range of the employed
FIR-wavelengths.
In addition stripe contact pairs with a large distance of 3~mm
were also used allowing irradiation of one contact only.

To allow polarization independent diffraction measurements at contact
edges, silver paint dots of 1~mm diameter
were applied on the edge of one metal
contact of the 3~mm slit sample covering partly the plain
superconductor surface (see Fig.~\ref{l3scan}).
The rough boundary of the silver paint spot yields
diffraction for all polarizations of the radiation field.

Measurements of the photoresponse were carried out using a TEA-CO$_2$ laser
pumped molecular FIR laser with NH$_3$ as active gas yielding laser lines of
$\lambda$ = 76, 90, 148 and 280~$\mu$m with a peak intensity up to
2~MW/cm$^2$
in 100~ns pulses.
The intensity of the laser pulses was controlled by
calibrated attenuators and monitored using a fast photon drag
p-Ge detector \footnote{ARTAS GmbH, model PD5F}.

The superconducting samples were biased by a fast constant
current source with bias current up to 100~mA and the
voltage across the sample in response to FIR laser pulses was
recorded with a digital storage oscilloscope. The measurements
were carried out in a temperature controllable
cryostat with optical access in the range 25 to 60~K.

\section{Results and Discussion}

\begin{figure}[h!]
\begin{center}           
            \mbox{\epsfxsize=9cm \epsfbox{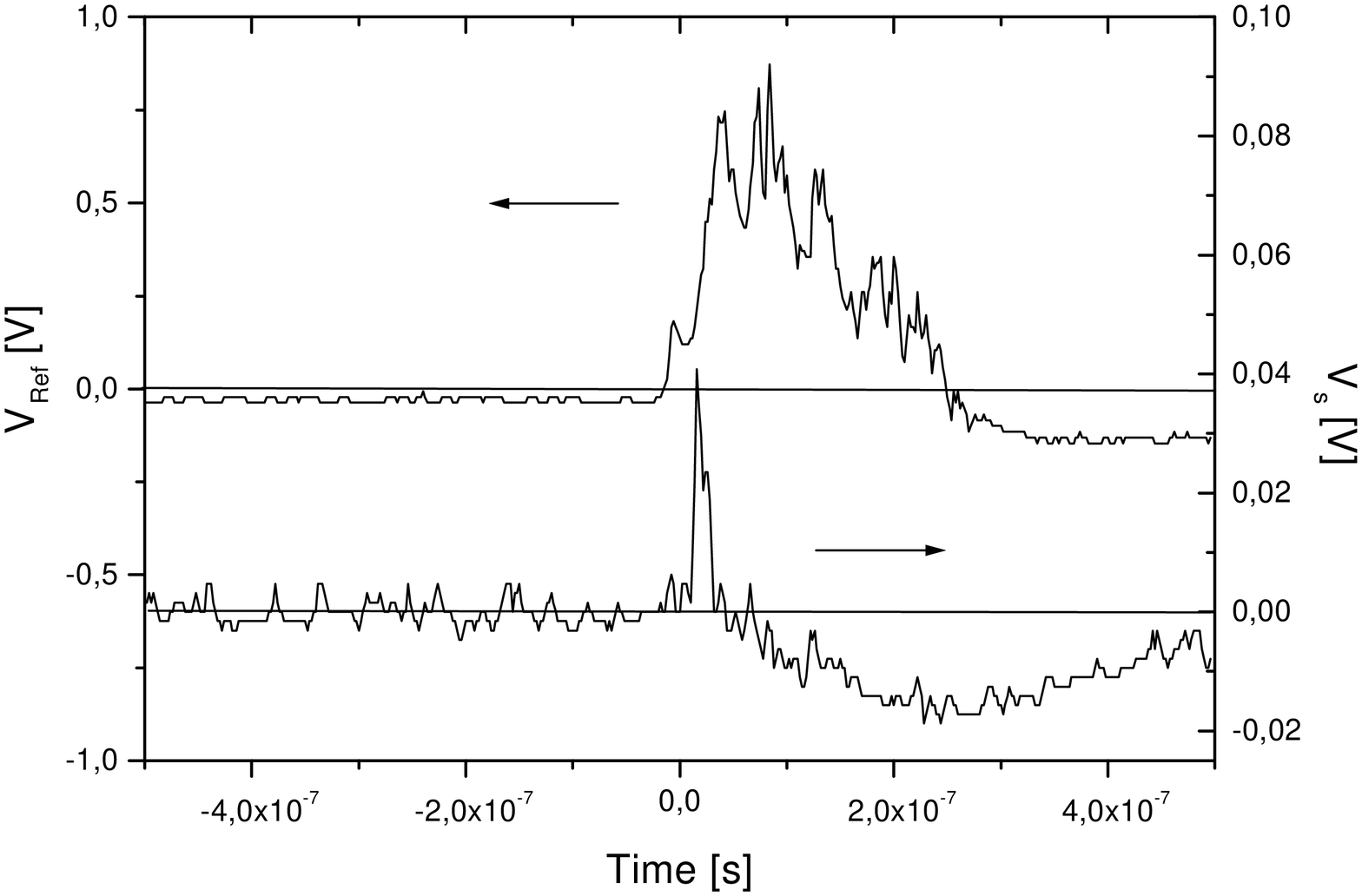}}
            
            \vspace{3mm}
            \mbox{\epsfxsize=9cm \epsfbox{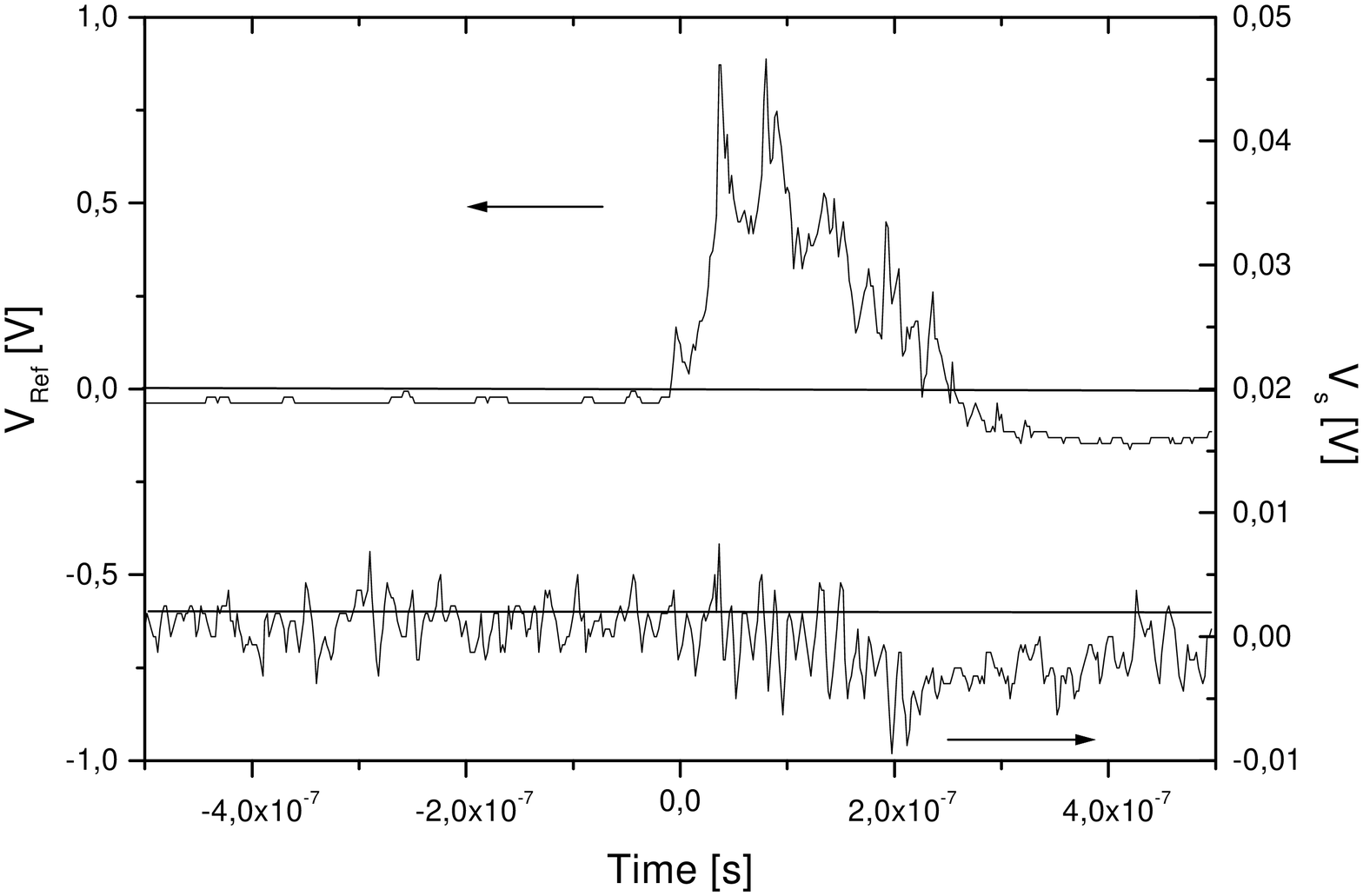}}
\end{center}            
\caption{\it{Typical signal pulse shapes compared to
reference photodetector signals, showing fast Josephson plus
bolometric signal at a contact edge (upper plate) and only bolometric 
response from the plain surface (lower plate).}}

\label{pulse}
\end{figure}

Fig.~\ref{pulse} shows the sample response to 76~$\mu$m
laser pulses (lower plates) in comparison to reference
detector signals (upper plates). The left lower plate shows
a signal trace when the laser was focused on the edge of the
the silver paint contact whereas in the lower right plate
the signal is shown which was observed when the laser was
focused on the superconductor away from the contacts. In the first
case  we see a fast positive signal ($\tau~<~10$~ns)
being synchronous to the leading edge of
the reference detector signal and a slower negative
signal ($\tau\approx 100$~ns). In the case when
solely the superconductor was irradiated only
the slow negative signal is found. Thus, the slow negative
signal can be identified with a bolometric effect whereas
the fast positive signal can be attributed to the Josephson
effect. In the signal trace bottom left the Josephson signal
is truncated by the bolometric signal of opposite sign.

\begin{figure}[h!]
\begin{center}
      \mbox{\epsfxsize=9cm \epsfbox{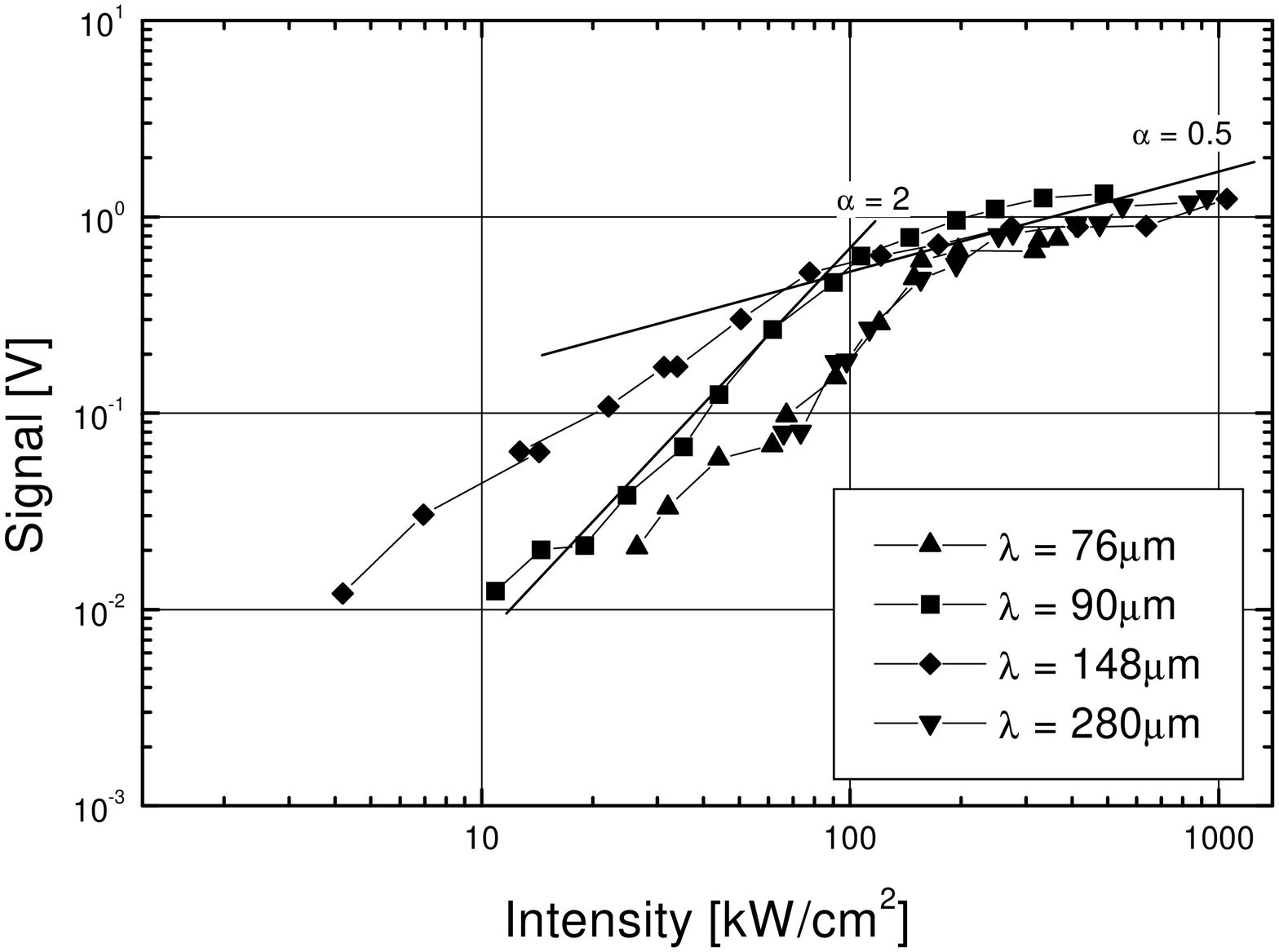}}

      \vspace{3mm}
      \mbox{\epsfxsize=9cm \epsfbox{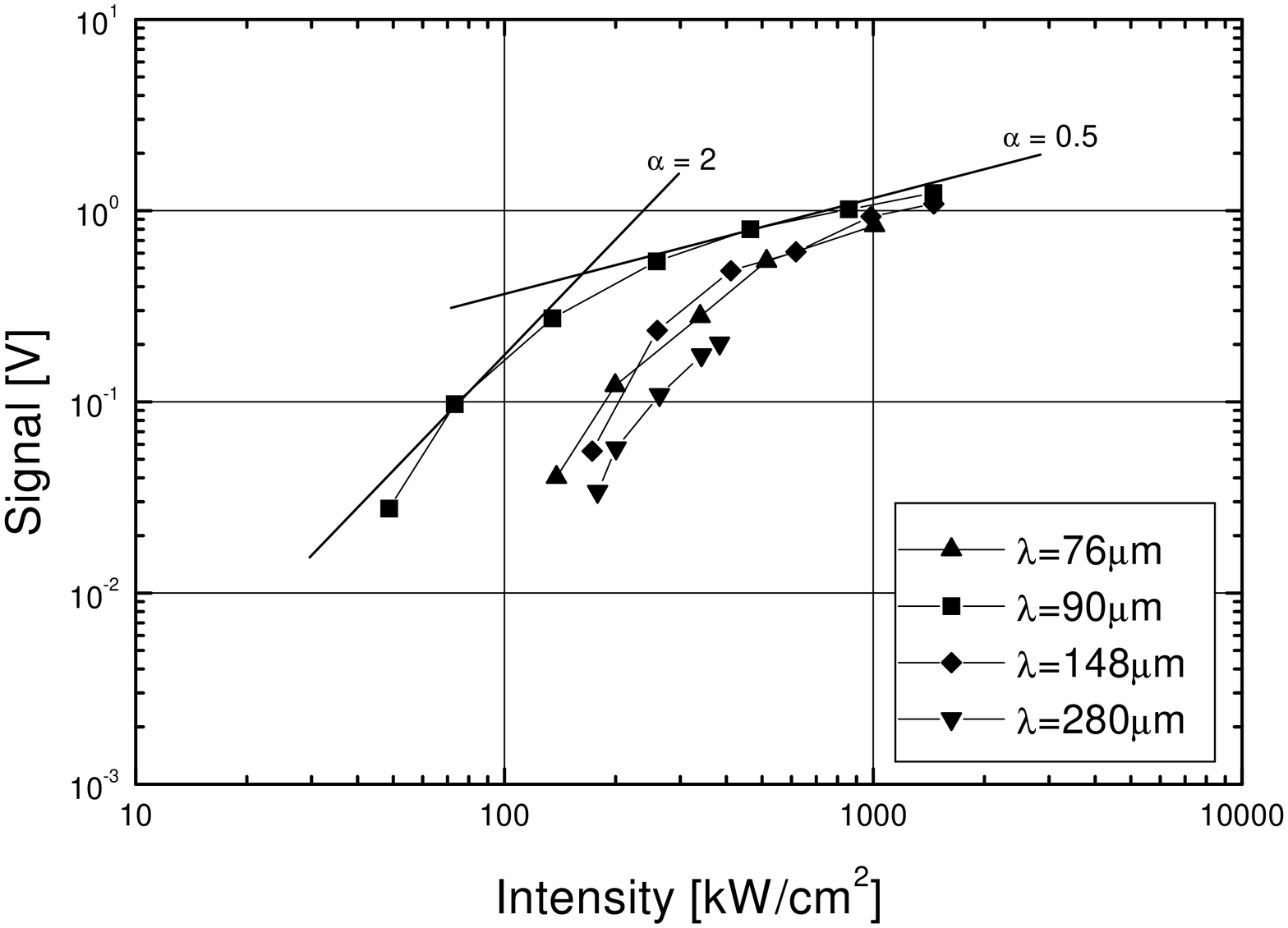}}
\end{center}
\caption{\it{FIR-intensity dependence of the Josephson signal
at a temperature of 40~K on the 3~mm~slit edge (upper plate) and on the
100~$\mu$m-slit (lower plate).}}

\label{l3intens}
\end{figure}

Fig. \ref{l3intens} shows the intensity dependence of the
signal voltage for different wavelengths demonstrating the typical
behavior of Josephson response following power
laws\cite{Huggard94Wavelength}. The left and right plates show the
response irradiating one contact edge of a 3~mm slit sample
and a 100~$\mu$m slit, respectively.
In both cases at low intensities an almost  wavelength independent
power law is observed
which proceeds at high intensities into signal $\propto
I^{\frac{1}{2}}$ where $I$ is the intensity.
The crossover occurs at practically the same intensity for all
wavelengths except for the 100~$\mu$m slit sample and
$\lambda = 90.5~\mu$m wavelength which is almost equal to
the contact distance.

\begin{figure}[h!]
\centerline{\mbox{\epsfxsize=9cm \epsfbox{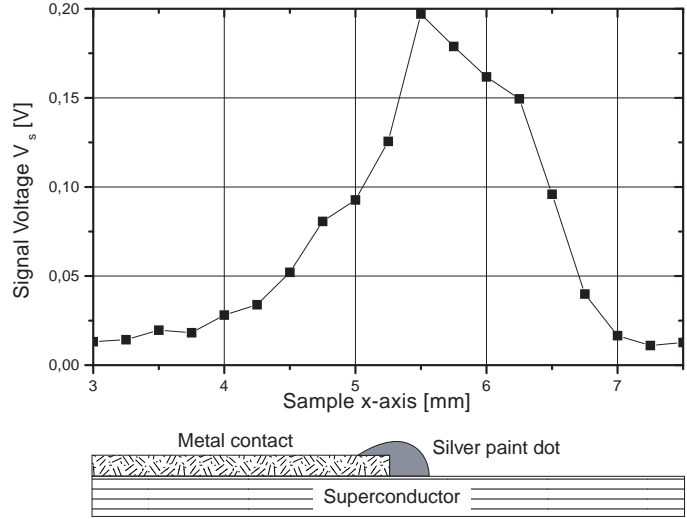}}}
\caption{\it{Signal height at scans of the laser focus across the
3~mm~slit.}}
\label{l3scan}
\end{figure}

In Fig.~\ref{l3scan} the signal voltage is shown
as a function of the spatial location of the laser focus scanned across the
sample surface perpendicular to the silver paint contact is.
The peak signal is plotted as a function of the center of
the focal spot with respect to the contact geometry.
The spatial signal distribution
correponds to the intensity profiles of the laser focus as
obtained by a high resolution pyroelectric
camera\footnote{Spiricon, Pyrocam I}.
A signal is observed only in the case of incidence onto the
silver paint contact edge and vanishes if only the plain
YBa$_2$Cu$_3$O$_{7-\delta}$
surface is irradiated. This observation gives evidence that
the signal must due to a component of the electric field of
the radiation which is normal to the c-axis oriented sample
modulating the critical current of the intrinsic Josephson
effect. Such a longitudinal electric field component may be
caused in the near-zone field of diffraction at the silver
paint contact.

The formation of effective intrinsic Josephson junctions in the c-axis of
YBa$_2$Cu$_3$O$_{7-\delta}$-layers is dependent of the oxygen
deficiency~$\delta$. A particular $\delta$ can be achieved by thermally
forced oxygen diffusion in and out of the sample along the ab-crystal planes.
Due to our sample geometry with large contact surfaces and its
c-axis orientation this diffusion is locally inhibited, leading to an
inhomogeneous oxygen distribution and therefore inhomogeneous
superconducting characteristics across the sample like different critical
temperatures and currents, as a semiconductor-like residual resistance in
the superconducting state. Electrical measurements also
indicate an inhomogeneous current distribution between the contacts, showing
several distinguished current paths.

\section{Conclusions}

In c-axis oriented thin oxygen depleted
YBa$_2$Cu$_3$O$_{7-\delta}$-films
a fast response to far infrared laser radiation has been
observed only if diffracting metallic structures on  top of
the film have been irradiated.
Therefore it is  concluded that the signal is due to suppression
of the critical current of the intrinsic Josephson coupling
between adjacent superconducting layers. This suppression
is caused by an
electric field component of the high frequency radiation
which is normal to the plane of the  superconducting film and
is generated in the near-zone field of diffraction.

\section{Acknowledgments}

Financial support by the Deutsche Forschungsgemeinschaft is
gratefully acknowledged. We thank H. Lengfellner for
advice in sample preparation and A.~Ya.\ Shulman, IRE
Moscow, for helpful discussions.

\end{document}